\documentclass[preprint,preprintnumbers,amsmath,amssymb]{revtex4}
\usepackage{graphicx}
\usepackage{bm}

\begin{document}

\title{Controlled multiqubit entangled states and quantum transmission in quantum molecule-spin systems}

\author{Xiang Hao}
\altaffiliation{Corresponding author:110523007@suda.edu.cn}

\author{Tao Pan}

\author{Taocheng Zang}

\author{Shiqun Zhu}

\affiliation{Department of Physics, School of Mathematics and
Physics, Suzhou University of Science and Technology, Suzhou,
Jiangsu 215011, People's Republic of China}

\begin{abstract}

The multiqubit entangled states are coherently controlled in the
quantum spin systems composed of $N+1$ interacting antiferromagnetic
molecular rings. The tunable intermolecular couplings arise from the
local exchange interactions between electron spins of $N$
circumjacent magnets and those of one central molecular ring. The
quantum dynamics of such system is analytically deduced by the
effective spin hamiltonian with anisotropic Heisenberg couplings. It
is found that entangled $W$ states can be generated with a high
precision under the circumstance of quantum fluctuations. The
multiqubit entangled state is also transferred from some molecular
rings to others by the selection of intermolecular couplings.

PACS: 03.67.Lx, 03.67.Mn, 75.50.Xx, 75.60.Jk
\end{abstract}

\maketitle

\section{Introduction}

Multiqubit entanglement is identified as a key resource for scalable
quantum computation and communication \cite{Nielsen00, Bose03}. As
an important multiqubit entangled state, an equally weighted
superposition of $N$ single-spin state, i.e., the entangled $W$
state, has also been applied to the quantum information processing
such as the Grover search \cite{Grover97}. In recent years, the
theoretical and experimental constructions of multiqubit entangled
states have been demonstrated with very few qubits in nuclear
magnetic resonance \cite{Chuang98, Ollerenshaw03} and trapped ions
\cite{Ivanov08}. However, this result is far away from the scalable
quantum information processing in principle. From this point of
view, significant efforts have been devoted to the study of quantum
dynamics of many spins in quantum dots \cite{Divin98},
semiconductors \cite{Awschalom02} and molecular magnets
\cite{Troiani05}. Electron spins in these solids are considered
among the promising candidates for quantum information technology
\cite{Zutic04}. Due to the range of neighboring spins in quantum
dots and semiconductors, the local control over the electrical and
magnetic field is very challenging \cite{Svore05}. This is becoming
one main technological obstacle in the implementation of quantum
logic gates. In this perspective, single molecular magnets(SMMs)
with antiferromagnetic interactions can act as effective qubits
\cite{Leuenberger01, Meier03}. Compared to quantum dots, the
molecular and supramolecular chemistry enable the manipulation of
long range interactions between coupled molecular rings
\cite{Bogania08}. Even for the fast control requirement, the
preferable electric control over SMMs is also possible
\cite{Trif08}. The proposal of quantum gates \cite{Carretta07} based
on two molecular magnets such as $Cr_{x}Ni(x=3,5,7)$ has been put
forward. In the interesting experiment \cite{Timco08}, $Cr_{x}Ni$
rings can be linked to each other so conveniently that the effective
coupling between them can be chemically tuned by choosing the
linker. Therefore, it is very valuable to investigate the multiqubit
entanglement in this quantum hardware which consists of a collection
of coupled molecular magnets.

In this paper, our attentions are focused on the special quantum
systems with $N$ antiferromagnetic molecular rings which are weakly
coupled to a central molecular magnetic. In Sec. II, for the typical
example of $Cr_{x}Ni$, the effective spin hamiltonian of $N+1$
coupled molecular magnets can be obtained in the form of anisotropic
Heiserberg exchange. In Sec. III, we analytically give the exact
quantum dynamics of the system in the single-spin subspace. The
entangled $W$ state is generated in this controllable quantum
systems. And the multiqubit entangled states can also be perfectly
transmitted from some rings to others by means of choosing the
intermolecular coupling. A useful discussion concludes the paper.

\section{Effective spin hamiltonian of coupled molecular magnets}

Some molecular magnets \cite{Choi06,Salman08} have been synthesized
as suitable candidates for carrying the quantum information and
qubit encoding. In particular, a class prototypical system of the
substituted antiferromagnetic $Cr_{x}Ni$ rings can act as an
effective two-level quantum system at low temperatures and show the
long decoherence time. Here, we take into account the quantum
systems of $Cr_{x_i}Ni$ rings with the number of $N$ which are
weakly coupled to a central molecular ring $Cr_{x_c}Ni$. The
hamiltonian of the quantum system can be described as
\begin{equation}
H=H_i+H_c+H_{int}
\end{equation}
Here $H_i$ and $H_c$ are the spin hamiltonians of the $i$th
circumjacent ring and central one respectively. According to
\cite{Affronte05}, the spin hamiltonian of $Cr_{x}Ni$ is
\begin{align}
H_x&=\sum_{k=1}^{x+1}J_{k}\vec{\tau}_{k}\cdot
\vec{\tau}_{k+1}+d_k[\tau_{k,z}^2-s_k(s_k+1)/3]& \nonumber\\
&+\sum_{k\neq l}\vec{\tau}_{k}\cdot \vec{D}_{k,l}\cdot
\vec{\tau}_{l},(x=x_{i},x_{c})&
\end{align}
where $\vec{\tau}_{k}$ is the spin operator of the $k$th electron
spin in a molecular ring, $k=x+1$ describes the site of $Ni$
substitution $(s_{x+1}=1)$ and others denote the sites of $Cr$ irons
$(s_k=3/2)$. The first item of the above equation is the dominant
isotropic Heisenberg exchange with different couplings
$J_{x+1}/J_k=a$, while the second and third ones account for the
anisotropic local crystal field and the intramolecular dipolar
interaction. For the experimental measurements, the intramolecular
dipolar coupling is small enough to be neglected in the following
context. With the idea of the local exchange between the $m$th
electron spin in the $i$th circumjacent ring and the $n$th one in
the central ring, the form of $H_{int}$ is expressed by
\begin{equation}
H_{int}=\sum_{i=1}^{N}\sum_{(m,n)}J_{m,n}^{i}\vec{\tau}_{m}^{i}\cdot
\vec{\tau}_{n}^{c}
\end{equation}
Here, the symbol of $\sum_{(m,n)}$ includes all possible pairs of
selective local linkers between two rings where $(m,n)$ denotes one
pair. At very low temperatures, all of the anitiferromagnetic
molecular rings serve as effective qubits for the ground doublet
states $\{|0_{i(c)}\rangle,|1_{i(c)}\rangle\}$ with the total spin
$S_{i(c)}=1/2$. These two degenerate states are separated from the
next excited state by a large energy gap $\Delta_g$. For the weak
local exchanges, the Heisenberg interaction of $H_{int}$ can be
expressed by the effective molecular spin operators $\vec{S}_{i(c)}$
\begin{equation}
H_{int}=\sum_{i=1}^{N}\frac
{\gamma_i}{2}[(S_{i}^{+}S_{c}^{-}+S_{i}^{-}S_{c}^{+})+(1+\Delta_i)S_{i}^{z}S_{c}^{z}]
\end{equation}
where the effective intermolecular interaction can be described by
$\gamma_i \thicksim \sum_{m,n}J_{m,n}^{i}\langle
1_{i}|\tau_{m}^{i,x}|0_{i}\rangle \cdot \langle
0_{c}|\tau_{n}^{c,x}|1_{c}\rangle$ and the anisotropy
$\Delta_i=1-\frac {\sum_{m,n}\langle
0_{i}|\tau_{m}^{i,z}|0_{i}\rangle \cdot \langle
0_{c}|\tau_{n}^{c,z}|0_{c}\rangle}{\sum_{m,n}\langle
1_{i}|\tau_{m}^{i,x}|0_{i}\rangle \cdot \langle
0_{c}|\tau_{n}^{c,x}|1_{c}\rangle}$. It shows that the interactions
closely depend on the definite structures of the molecular rings and
local linkers. The recent experiment has demonstrated that the
effective couplings can be chemically tuned by the control of the
local linkers between two molecular rings \cite{Timco09}. Without
losing the generality, the controllable anisotropy is calculated for
two coupling rings $Cr_{3}Ni$ with the same structure. It is seen
that the anisotropy $\Delta$ can be varied with the change of the
ratio of the intramolecular couplings $a$ and local crystal field
$d$ in Fig. 1(a). When the value of $a$ is near to one, the values
of $\Delta$ is increased almost linearly with respect to $b$. By the
manipulation of two pairs of local linkers, Fig. 1(b) illustrates
that the anisotropy $\Delta$ can also be tunable in the large range
from the negative value to the positive one. For the special case of
$\Delta=-1$, the model of $H_{int}$ is simplified to be the
Heisenberg $XX$ one. If the selective local exchanges arrive at a
critical ones $b=b_c$, the anisotropy can be rapidly tuned from one
negative value to a large positive one. Because of the symmetric
property $[H_{int},\sum_{i=1}^{N}S_{i}^{z}+S_{c}^{z}]=0$, the
effective spin Hamiltonian for $N+1$ interacting molecular rings can
be described in the single-spin subspace of
$\{|\psi_1\rangle,|\psi_2\rangle,\cdots,|\psi_N\rangle,|\psi_{N+1}\rangle
\}$
\begin{align}
H_{eff}|\psi_i\rangle&=\frac
{\gamma_i}4[(N-2)(1+\Delta_i)|\psi_i\rangle+2|\psi_{N+1}\rangle]& \nonumber\\
H_{eff}|\psi_{N+1}\rangle&=\sum_{i=1}^{N}\frac
{\gamma_i}4[2|\psi_i\rangle-(1+\Delta_i)|\psi_{N+1}\rangle]&
\end{align}
where the basis of the subspace $|\psi_i\rangle=|1_i\rangle \otimes
\prod_{k\neq i}|0_k\rangle$. At very low temperatures, the quantum
information could be processed in this subspace. Then, the dynamics
of these states can be analytically solved as follows.

\section{Generation and transmission of multiqubit entangled states}

Our quantum memory registers will be expanded by the set of states
in the single-spin subspace for low temperatures $k_BT\leq
\Delta_g$. The general expression of the effective Hamiltonian
$H_{eff}$ can be obtained
\begin{widetext}
\begin{equation}
H_{eff}=\frac {1}{4}\left(\begin{array}{ccccc}
\gamma_1(1+\Delta_1)(N-2) & 0 & \cdots & 0 & 2\gamma_1 \\
0 & \gamma_2(1+\Delta_2)(N-2) & \cdots & 0 & 2\gamma_2 \\
\vdots & \vdots & \ddots& \vdots & \vdots\\
0 & 0 & \cdots & \gamma_N(1+\Delta_N)(N-2) & 2\gamma_N \\
2\gamma_1 & 2\gamma_2 & \cdots & 2\gamma_N &
-\sum_{i}^{N}\gamma_i(1+\Delta_i)
\end{array}\right)
\end{equation}
\end{widetext}
By means of the tunable exchanges $J_{m,n}^{i}$ for the local
selective linkers, the anisotropic interactions can be controlled
and satisfy $\gamma_i(1+\Delta_i)=C$. In this condition, we can
analytically obtain the eigenstates and corresponding eigenvalues .
The effective Hamiltonian $H_{eff}$ has $N-1$ degenerate states
$|\phi_i\rangle$ with the same eigenvalue $\lambda_i=\frac
{C(N-2)}4$ and two nondegerate states $|\phi_j\rangle(j=N,N+1)$ with
$\lambda_{N,N+1}=\frac 14[-C\pm\sqrt{4\Omega^2+C^2(N-1)^2}]$ where
$\Omega=\sqrt{\sum_i^N\gamma_i^2}$ . The set of all orthonormalized
eigenstates can be given by
\begin{align}
|\phi_i\rangle&=\frac
{1}{\Gamma_i\Gamma_{i+1}}(\gamma_1\gamma_{i+1},\cdots,\gamma_i\gamma_{i+1},-\Gamma_i^2,0,\cdots,0)^{T}&\nonumber
\\
|\phi_j\rangle&=\frac {X_j}{\Gamma_N}(\gamma_1,\cdots,\gamma_N,Y_j\Gamma_N/X_j)^{T},(j=N,N+1)&\nonumber \\
\end{align}
where the parameters $\Gamma_i=\frac
{2\sqrt{\sum_k^{i}\gamma_k^2}}{C(N-2)}$ and $X_j=\frac
1{\sqrt{1+(\lambda_j-1)^2/\Gamma_N^2}},Y_j/X_j=\frac
{\lambda_j-1}{\Gamma_j}(j=N,N+1)$. To study how to generate the
multiqubit entangled states, we need to find the evolution operator
in the single-spin subspace
\begin{equation}
U(t)=\sum_k^{N+1}\Lambda_k(t)|\phi_k\rangle\langle\phi_k|
\end{equation}
where $\Lambda_k=\Lambda=\exp{(-i\int_0^t \frac {C(N-2)}4
d\tau}),(k\neq N,N+1)$ and $\Lambda_{N,N+1}=\exp{(-i\int_0^t
\lambda_{N,N+1} d\tau})$.

For illustration, we choose two kind of typical initial states. If
the initialized state is $|\Psi(0)\rangle=|\psi_i\rangle_{i\neq
N+1}$, the general state at any time $t$ is obtained by
\begin{align}
|\Psi(t)\rangle=&\sum_{m\neq i}\frac
{\gamma_m\gamma_i}{\Omega^2}(R-\Lambda)|\psi_m\rangle&
\nonumber \\
&+\frac {\gamma_i}{\Omega}S|\psi_{N+1}\rangle +[\Lambda-\frac
{\gamma_i^2}{\Omega^2}(\Lambda-R)]|\psi_i\rangle&
\end{align}
Here $R=\frac {\Lambda}{1+A^2}(e^{i\theta_1}+A^2e^{-i\theta_2})$,
$S=-\frac {A\Lambda}{1+A^2}(e^{i\theta_1}-e^{-i\theta_2})$ and
$A=\frac {C(N-1)}{2\Omega}-\sqrt{1+\frac {C^2(N-1)^2}{4\Omega^2}}$
where the phase angles $\theta_1=\int_0^{t}\frac {\Omega A}{2}d\tau$
and $\theta_2=\int_0^{t}\frac {\Omega }{2A}d\tau$. At a certain time
for $\theta_1+\theta_2=\int\frac {\Omega
(A^2+1)}{2A}=\pm2k\pi(k=0,1,2,\ldots)$, the possibility of
$|\psi_{N+1}\rangle$ is zero. For the circumjacent $N$ sites, we
need the possibility of each $|\psi_m\rangle_{m\neq i}$ is equal and
the ratio of the effective interactions $p= \gamma_m^2/\gamma_i^2$
\begin{equation}
p=\frac {1-N\cos\theta_1\pm \sqrt{2N(1-\cos \theta_1)-N^2\sin^2
\theta_1}}{(N-1)^2}
\end{equation}
The actual quantum state at this time is $|\Psi\rangle=\frac
{1}{\sqrt{N}}(\sum_{m\neq
i}|\psi_m\rangle+e^{i\chi}|\psi_i\rangle$). Then by means of the
single-qubit phase operator at the $i$-th ring, we can obtain the
perfect $W$ state. When another initial state is chosen as
$|\psi_{N+1}\rangle$, the general state at time $t$ is expressed by
\begin{equation}
|\Psi\rangle=\sum_{m}\frac {\gamma_m}{\Omega}S|\psi_m\rangle+\frac
{\Lambda(A^2e^{i\theta_1}+e^{-i\theta_2})}{1+A^2}|\psi_{N+1}\rangle
\end{equation}
Through the calculation of the possibilities of the eigenstates, it
is found out that the entangled $W$ state $|\Psi_W\rangle=\frac
{1}{\sqrt{N}}\sum_m|\psi_m\rangle$ can be generated when the time
satisfies $\theta_1+\theta_2=\pm(2k+1)\pi$, $\Delta=-1$ and
$\gamma_m=\gamma$. However, in realistic quantum control, certain
fluctuations from internal and external impacts are unavoidable. To
evaluate the effects of quantum fluctuation on the generation of the
multiqubit entangled state, we take into account the simplest case
of $N=3$ under the circumstance of $|\Psi(0)\rangle=|\psi_3\rangle$
and $\gamma_1=\gamma_2$. In accordance with the above analysis, the
perfect entangled $W$ state can be produced at the time
$t_W=2|k|\pi/\sqrt{C^2+\Omega^2}$. In regard to the certain
fluctuation, the effective intermolecular interactions possibly obey
the relation of $\gamma_3(1+\Delta_3)=C(1+\delta)\neq
\gamma_{1(2)}(1+\Delta_{1(2)})=C$. From Fig. 2, it is seen that the
error of generation $E_r=1-|\langle \Psi_W|\Psi(t_W) \rangle|^2$ is
almost linearly decreased with the fluctuation parameter $\delta$.
The relation of the values $E_r\sim |\delta|/ 10$ can demonstrate
that the proficiency of the generation of the multiqubit entangled
state is high enough to resist the fluctuation to a certain extant.

In solid state quantum computers, it is very necessary to
investigate the quantum transport of the multiqubit states. Here we
provide a efficient scheme of transferring $L$-qubit entangled
states in our quantum memory registers. For low temperatures, the
initial $L$-qubit entangled state for $L \leq \frac {N-1}2$ can be
expanded in the single-spin subspace
$|\Phi(0)\rangle=\sum_{i=1}^{L}c_i|\psi_i\rangle,(\sum|c_i^2|=1)$.
For the perfect quantum transmission, the entangled state can occur
at another $L$ sites from the $L+1$-th ring to $2L$-th one after the
time. By the analytical calculation, it is found out that the
perfect quantum transfer can be achieved in the condition of
\begin{equation}
\frac {\gamma_i}{\gamma_j}=\frac {c_i}{c_j}=\frac
{\gamma_{L+i}}{\gamma_{L+j}},\gamma_{m>2L}=0
\end{equation}
This means that the $L$-qubit entangled state can be transported
perfectly by the adjustment of the effective intermolecular
interactions. For an example of $N=5$, the transferred state is
$|\Phi(0)\rangle=\sin \alpha |\psi_1\rangle+\cos \alpha
|\psi_2\rangle$ where the two-qubit entangled state $(\sin \alpha
|10\rangle_{1,2}+\cos \alpha |01\rangle_{1,2})$ exists in two rings
of $i=1,2$. To access the quantum transfer in this system, the
fidelity at time $t$ is used in the form of
\begin{equation}
F=|\langle \Phi(t)|\Phi(0)\rangle |^2
\end{equation}
It is seen that the values of fidelity take on the periodic
evolution with the time in Fig. 3. The perfect quantum transmission
can be realized by the efficient manipulation of the effective
intermolecular couplings.

\section{Discussion}

The multiqubit entangled $W$ state can be generated by chemically
tuning the effective intermolecular interactions in the quantum
system of $N+1$ weakly coupled antiferromagnetic molecular rings.
This method can provide an important entanglement source for the
scalable quantum search in solids. In the present work, the
anisotropy of the effective spin Hamiltonian can be chemically
controlled by the selection of local linkers between electron spins
of different rings. For low temperatures, the state carrying quantum
information can be expanded in the single-spin subspace. The high
precision of producing $W$ state at $N$ circumjacent rings can be
satisfied with the consideration of a certain fluctuations. And the
$L$-qubit entangled state can also be perfectly transferred at a
series of periodic time in our quantum registers by means of the
control of the effective long range couplings between molecular
rings. This is also an efficient proposal for the quantum router
\cite{Bose05} in which the communication can be directed between any
chosen rings.

\section{Acknowledgement}

This work was supported by the Initial Project of Research in SUST
and the National Natural Science Foundation of China No. 10774108.

\newpage

{\Large Figure caption}

Figure 1

(a). The anisotropy $\Delta$ is plotted as functions of the ratio of
the intramolecular couplings $a$ and local crystal field $d$ for
$J_k=17$; (b). The anisotropy $\Delta$ is tuned by the control of
two pairs of linkers $b=J_{4,4}/J_{1,2}$ for $J_{1,2}=1$,
$J_{k}=17$, $a=0.9$ and $d=0.3$.

Figure 2

The error $E_r$ for the generation of $W$ state is plotted as a
function of the fluctuation $\delta$ for $\gamma_1=\gamma_2=1$.

Figure 3

The fidelity of the quantum state transfer is plotted with the time
for the case of $\alpha=\pi/4$,
$\gamma_1/\gamma_2=\gamma_3/\gamma_4=1$ and $\gamma_5=0$.


\begin{references}
\bibitem{Nielsen00} M. A. Nielsen and I. L. Chuang, {\it Quantum
Computation and Quantum Information}(Cambridge University Press,
Cambridge, 2000).
\bibitem{Bose03} S. Bose, Phys. Rev. Lett. \textbf{91},
207901(2003).
\bibitem{Grover97} L. K. Grover, Phys. Rev.
Lett. \textbf{79}, 325(1997).
\bibitem{Chuang98} I. L. Chuang, N. Gershenfeld and M. Kubinec,
Phys. Rev. Lett. \textbf{80}, 3408(1998).
\bibitem{Ollerenshaw03} J. E. Ollerenshaw, D. A. Lidar and L. E.
Kay, Phys. Rev. Lett. \textbf{91}, 217904(2003).
\bibitem{Ivanov08} S. S. Ivanov, P. A. Ivanov and N. V. Vitanov,
Phys. Rev. A \textbf{78}, 030301(R)(2008).
\bibitem{Divin98} D. Loss and D. P. DiVincenzo, Phys. Rev. A \textbf{57},
120(1998).
\bibitem{Awschalom02} D. Awschalom, N. Samarsh and D. Loss, {\it Semiconductor Spintronics
and Quantum Computation}(Springer, Berlin, 2002).
\bibitem{Troiani05} F. Troiani, A. Ghirri, M. Affronte, S. Carretta,
P. Santini, G. Amoretti, S. Piligkos, G. Timco and R. E. P.
Winpenny, Phys. Rev. Lett. \textbf{94}, 207208(2005).
\bibitem{Zutic04} I. \v{Z}utic, J. Fabian and S. Das Sarma, Rev. Mod.
Phys. \textbf{76}, 323(2004).
\bibitem{Svore05} K. M. Svore, B. M. Terhal and D. P. DiVincenzo,
Phys. Rev. A \textbf{72}, 022317(2005).
\bibitem{Leuenberger01} M. Leuenberger and D. Loss, Nature(London)
\textbf{410}, 789(2001).
\bibitem{Meier03} F. Meier, J. Levy and D. Loss, Phys. Rev.
Lett. \textbf{90}, 047901(2003).
\bibitem{Bogania08} L. Bogania and W. Wernsdorfer, Nature
Mater. \textbf{7}, 179(2008).
\bibitem{Trif08} M. Trif, F. Troiani, D. Stepanenko and D. Loss,
Phys. Rev. Lett. \textbf{101}, 217201(2008).
\bibitem{Carretta07} S. Carretta, P. Santini, G. Amoretti, F. Troiani and M. Affronte, Phys.
Rev. B. \textbf{76}, 024408(2007).
\bibitem{Timco08}G. A. Timco, S. Carretta, F. Troiani, F. Tuna, R. J. Pritchard, C.r A. Muryn, E. J. L. McInnes, A. Ghirri, A. Candini and P.
Santini, Nature Nanotechnology \textbf{4}, 173(2008)
\bibitem{Choi06} K. Y. Choi, Y. H. Matsuda, H. Nojiri, U. Kortz, F. Hussain, A. C. Stowe, C. Ramsey and N. S. Dalal, Phys. Rev. Lett. \textbf{96},
107202(2006).
\bibitem{Salman08} Z. Salman, R. F. Kiefl, K. H. Chow, W. A.
MacFarlane, T. A. Keeler, T. J. Parolin, S. Tabbara and D. Wang,
Phys. Rev. B \textbf{77}, 214415(2008).
\bibitem{Affronte05} F. Troiani, M. Affronte, S. Carretta,
P. Santini and G. Amoretti, Phys. Rev. Lett. \textbf{94},
190501(2005).
\bibitem{Bose05} S. Bose, B. Q. Jin and V. E. Korepin, Phys. Rev.
A \textbf{72}, 022345(2005).

\end{references}
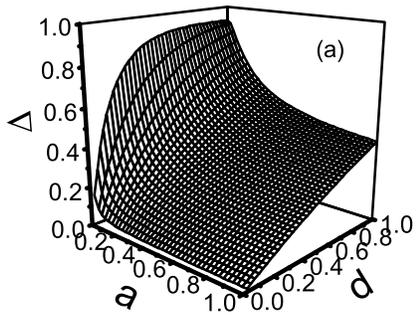
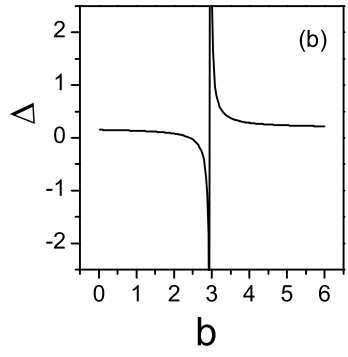
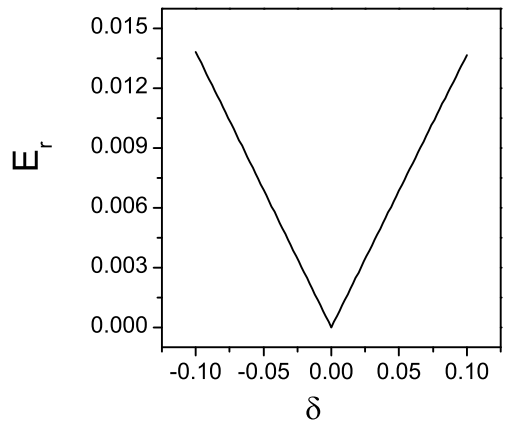
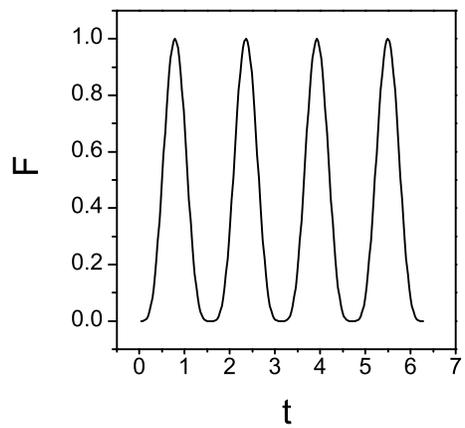
\end{document}